\documentclass[letterpaper, 10 pt, conference]{ieeeconf}  

\IEEEoverridecommandlockouts                              

\usepackage{graphics} 
\usepackage{epsfig} 
\usepackage{mathptmx} 
\usepackage{times} 
\usepackage{amsmath} 
\usepackage{amssymb}  
\usepackage{cite}
\usepackage{xcolor}
\usepackage{soul}
\usepackage{hyperref}
\usepackage{fancyhdr}

\fancypagestyle{plain}{
  \fancyhf{}
  \fancyhead[C]{Conference on \LaTeX}     
  \fancyfoot[L]{This is a notice}

}
\usepackage{eso-pic}

\title{\LARGE \bf
 Towards Non-contact 3D Ultrasound for Wrist Imaging}

\author{Antony Jerald$^{1}$, Madhavanunni A. N.$^{1}$, Gayathri Malamal$^{1}$ and Mahesh Raveendranatha Panicker$^{1}$
\thanks{$^{1}$Antony Jerald, Madhavanunni A. N., Gayathri Malamal and Mahesh Raveendranatha Panicker (Email: mahesh@iitpkd.ac.in) are with the Centre for Computational Imaging and Department of Electrical Engineering, Indian Institute of Technology Palakkad, Kerala, India}}

\begin{document}

\AddToShipoutPictureBG*{%
  \AtPageUpperLeft{%
    \setlength\unitlength{1in}%
    \hspace*{\dimexpr0.5\paperwidth\relax}
    \makebox(0,-0.75)[c]{\textcolor{red}{\large This is an originally submitted version and has not been reviewed by independent peers.}}%
}}

\AddToShipoutPictureBG*{%
  \AtPageLowerLeft{%
    \setlength\unitlength{1in}%
    \hspace*{\dimexpr0.5\paperwidth\relax}
    \makebox(0,0.75)[c]{\textcolor{red}{\textit{This work is licensed under a \href{https://creativecommons.org/licenses/by-nc-nd/4.0/}{Creative Commons Attribution-NonCommercial-NoDerivatives (CC-BY-NC-ND) 4.0 License.}}}}%
}}

\maketitle
\thispagestyle{empty}
\pagestyle{empty}

\begin{abstract}
\textit{Objective:} The objective of this work is an attempt towards non-contact freehand 3D ultrasound imaging with minimal complexity added to the existing point of care ultrasound (POCUS) systems. \textit{Methods:} This study proposes a novel approach of using a mechanical track for non-contact ultrasound (US) scanning.  The approach thus restricts the probe motion to a linear plane, to simplify the acquisition and 3D reconstruction process.  A pipeline for US 3D volume reconstruction employing an US research platform and a GPU-based edge device is developed. \textit{Results:}  The efficacy of the proposed approach is demonstrated through \textit{ex-vivo} and \textit{in-vivo} experiments. A MATLAB application of the proposed non-contact 3D reconstruction and sample data acquired are available at the \href{https://github.com/antony333/Towards-Noncontact-3D-Ultrasound-for-Wrist-Imaging/tree/main#readme}{link}. \textit{Conclusion:} The proposed approach with the adjustable field of view capability, non-contact design, and low cost of deployment without significantly altering the existing setup would open doors for up gradation of traditional systems to a wide range of 3D US imaging applications. \textit{Significance:} Ultrasound (US) imaging is a popular clinical imaging modality for the point-of-care bedside imaging, particularly of the wrist/knee in the pediatric population due to its non-invasive and radiation free nature. However, the limited views of tissue structures obtained with 2D US in such scenarios make the diagnosis challenging. To overcome this, 3D US imaging which uses 2D US images and their orientation/position to reconstruct 3D volumes was developed. The accurate position estimation of the US probe at low cost has always stood as a challenging task in 3D reconstruction. Additionally, US imaging involves contact, which causes difficulty to pediatric subjects while monitoring live fractures or open wounds. Towards overcoming these challenges, a novel framework is attempted in this work.
\end{abstract}

\section{Introduction}
\label{sec:introduction}
Ultrasound imaging is a popular point-of-care medical imaging modality due to its advantages such as dynamic imaging, lack of ionizing radiation, comparatively low cost, and easy disinfection. It typically follows the pulse-echo approach, which involves sending acoustic signals into the subject's body and receiving the reflections from the tissue interfaces using an array of transducer elements. The recorded reflections are beamformed to reconstruct the B-mode (brightness mode) ultrasound image. However, in 2D ultrasound, the clinician needs to mentally transform a sequence of 2D images to produce a 3D volume representation to arrive at a diagnosis. It is also challenging to relocate the exact orientation of a previously captured image to record the progression and regression of pathology in response to therapy which increases the inaccuracies and procedure times  \cite{c1}. Further, conventional ultrasound imaging involves direct contact of the transducer with the subject, where a thin layer of ultrasound gel is commonly employed as a couplant between the transducer and the subject's body. This leads to contact-sensitive images as the probe manipulation is clinician dependent and the variable pressure applied on the probe may result in tissue compression \cite{c30}. Also, the pressure when applied on delicate pediatric wrists or injured/wounded areas, brings discomfort to the subjects. In this regard, there is a growing need for non-contact 3D ultrasound imaging, to obtain more precise and integrated tissue information to improve the clinical outcomes \cite{c14} by ensuring comfort to the subject, which is the focus of this work.

\subsection{Related Works}
\label{sec:Related Works}
3D ultrasound imaging typically involves (unless employing a very expensive and bulky 2-D matrix array probe) stacking 2D ultrasound frames from different positions and angles and interpolating them to generate a 3D volume. The three main stages in 3D ultrasound reconstruction are data acquisition, volume reconstruction, and rendering. Acquisition refers to collecting the 2D ultrasound images and their positions and orientations. The reconstruction process involves the interpolation and approximation algorithms to fit these acquired images into a regular volume grid with a higher resolution. The rendering process displays the volume data from the voxel array for clinical analysis.

Various ultrasound data acquisition methods using 2D arrays and mechanical 3D probes have been developed over the years. Using a dedicated 2D array (matrix) probe is one of the fastest ways to view the 3D volume in real-time as it does not require precise orientation information for volume generation \cite{c16, c50}. But a 2D array probe is expensive and complex to design in terms of hardware and software. Moreover, the limited size of such probes, results in a smaller field of view in imaging \cite{c1}. The mechanical 3D probes are also popular, where a linear array transducer is motorized to translate, rotate and tilt within the probe to acquire images. However, the mechanical 3D probes must be held statically by the clinicians which introduces latent flaws in image acquisition \cite{c2}. Another method uses motorized mechanisms to tilt or translate the conventional linear array probe and rapidly acquire 2D ultrasound images \cite{c52,c18}. Here, the scanning position and orientation need to be predefined and are controlled by a stepper/servo motor, hence acquiring regularly spaced 2D ultrasound frames. However, such probes are bulky making them inconvenient for frequent scanning \cite{c2}. A 3D ultrasound reconstruction is also achieved with freehand acquisition, where orientation/position estimation sensors like electromagnetic (EM) tracking sensors or optical sensors register the orientation/position of each frame along with the image. EM tracker observes electrical currents that are generated by a receiver attached to the probe which moves within a magnetic field created by either an alternating current or a direct current transmitter. This method allows the user to move the probe freely, increasing its movement flexibility  \cite{c3}. However, EM trackers can be affected by interference from nearby magnetic sources, decreasing tracking accuracy \cite{c1}. Optical tracking methods use two or three integrated cameras to track infrared light-emitting or reflective markers attached to the probe \cite{c45,c44} and the tracking accuracy is improved by increasing the number of cameras or markers. However, employing additional cameras may result in calibration errors when forming a stereo view. Moreover, the high cost of these position-tracking devices limits their application in clinical settings. Sensorless freehand data acquisition methods have also emerged, which use information from ultrasound images to determine the probe's orientation. Such methods include deep learning-based sensorless tracking \cite{c40,c41,c48} and speckle decorrelation methods \cite{c47,c49,c42}. Even though sensorless methods are convenient, the inconsistency in scan rates and angles can lead to non-smooth reconstruction and lower the image quality during visualization \cite{c43}. Nevertheless, deep learning-based methods still require EM or optical trackers to generate training data.

Volume reconstruction methods are employed for 3D reconstruction from the 2D ultrasound images. The common volume reconstruction methods include, the voxel-based methods (VBMs), pixel-based methods (PBMs), and function-based methods (FBMs) \cite{c35}. VBMs involve traversing each voxel in the target voxel grid and extracting information from the input 2D images to assign values to the voxels. A commonly used algorithm is the voxel nearest neighbor algorithm (VNN), which scans each voxel in the volume and selects the nearest pixel value from a set of 2D frames that is to be assigned to the voxel \cite{c36}. While this approach effectively preserves the original texture patterns from the 2D frames, it can cause significant reconstruction artefacts when there is a considerable distance between the voxel and the corresponding 2D frames \cite{c1}. The PBMs involve traversing each pixel in the input images and assigning the pixel value to one or multiple voxels. A commonly used algorithm within the pixel-based method is the pixel nearest neighbor (PNN). It iterates through each pixel in the 2D images and assigns each pixel value to the nearest voxel. When multiple contributions exist for the same voxel, pixels are typically averaged to determine the final value. Empty voxels are filled by utilizing the average, maximum, minimum, or median of the neighbouring voxel values which are already assigned \cite{c37}. The PNN algorithm causes blurred results when too many empty voxels are present in the reconstructed volume \cite{c1}. 
The functional basis methods (FBMs) implement functional interpolation for 3D ultrasound reconstruction and use pixel values and relative positions to determine the function coefficients \cite{c60,c61}. The functions are evaluated at regular intervals to generate the voxel array. The approach enables the reconstruction of a 3D image with improved accuracy but is computationally expensive \cite{c38}.

The rendering of the reconstructed 3D volume is typically achieved through multiplanar reformatting, volume, and surface rendering. In multiplanar reformatting, resliced 2D ultrasound planes are extracted from the 3D volume and are displayed with a 3D impression \cite{c62,c1} providing three orthogonal slice views: transverse, coronal, and sagittal planes. The volume rendering utilizes ray-casting or ray-marching techniques to project changes in light passing through the 3D volume, enabling comprehensive  3D visualization \cite{ c64, c63}. Surface rendering generates a 3D surface based on the segmented boundary data points obtained by creating surface triangles or polygons created through interpolation \cite{c64}.

\subsection{Contributions}
Most approaches in literature enable 3D imaging on a moderately high-cost system requiring good computational capacity and are not intended for point-of-care imaging. In this work, a novel non-contact approach to overcome the challenge of orientation/position estimation in freehand-based scanning for wrist imaging is attempted. A preliminary work in this direction has been proposed by us in \cite{confpaper}, where the initial ex-vivo and phantom results are presented. In this work, the full design aspects, through experimental analysis and also a framework for the 3D reconstruction pipe-line and protocol with emphasis on non-contact scanning are presented. The proposed 3D reconstruction pipeline utilizes a cost-effective mechanical track and high frame rate plane wave imaging to streamline probe tracking and volume reconstruction. By enabling linear planar motion of the probe and with water as a couplant, the mechanical track is particularly well-suited for applications like non-contact wrist imaging. The approach is cheaper and offers greater convenience by eliminating the need for specialized transducers for imaging. Additionally, it addresses the challenges of tissue compression and subject discomfort associated with conventional ultrasound imaging.

\AddToShipoutPictureBG*{%
  \AtPageUpperLeft{%
    \setlength\unitlength{1in}%
    \hspace*{\dimexpr0.5\paperwidth\relax}
    \makebox(0,-0.75)[c]{\textcolor{red}{\large This is an originally submitted version and has not been reviewed by independent peers.}}%
}}

\AddToShipoutPictureBG*{%
  \AtPageLowerLeft{%
    \setlength\unitlength{1in}%
    \hspace*{\dimexpr0.5\paperwidth\relax}
    \makebox(0,0.75)[c]{\textcolor{red}{\textit{This work is licensed under a \href{https://creativecommons.org/licenses/by-nc-nd/4.0/}{Creative Commons Attribution-NonCommercial-NoDerivatives (CC-BY-NC-ND) 4.0 License.}}}}%
}}

\section{Proposed Approach}
\label{Proposed Approach}
The proposed approach consists of a 3D printed low-cost mechanical track for tilt-resistant linear scans, an ArUco marker-based probe tracking setup alongside a high frame rate ultrasound imaging with a linear array probe, and a low complex 3D interpolation and rendering pipeline for volume segmentation/visualization.

\subsection{Design of the Mechanical Track}

\begin{figure*}[t]
    \centering
    \includegraphics[width=1\textwidth]{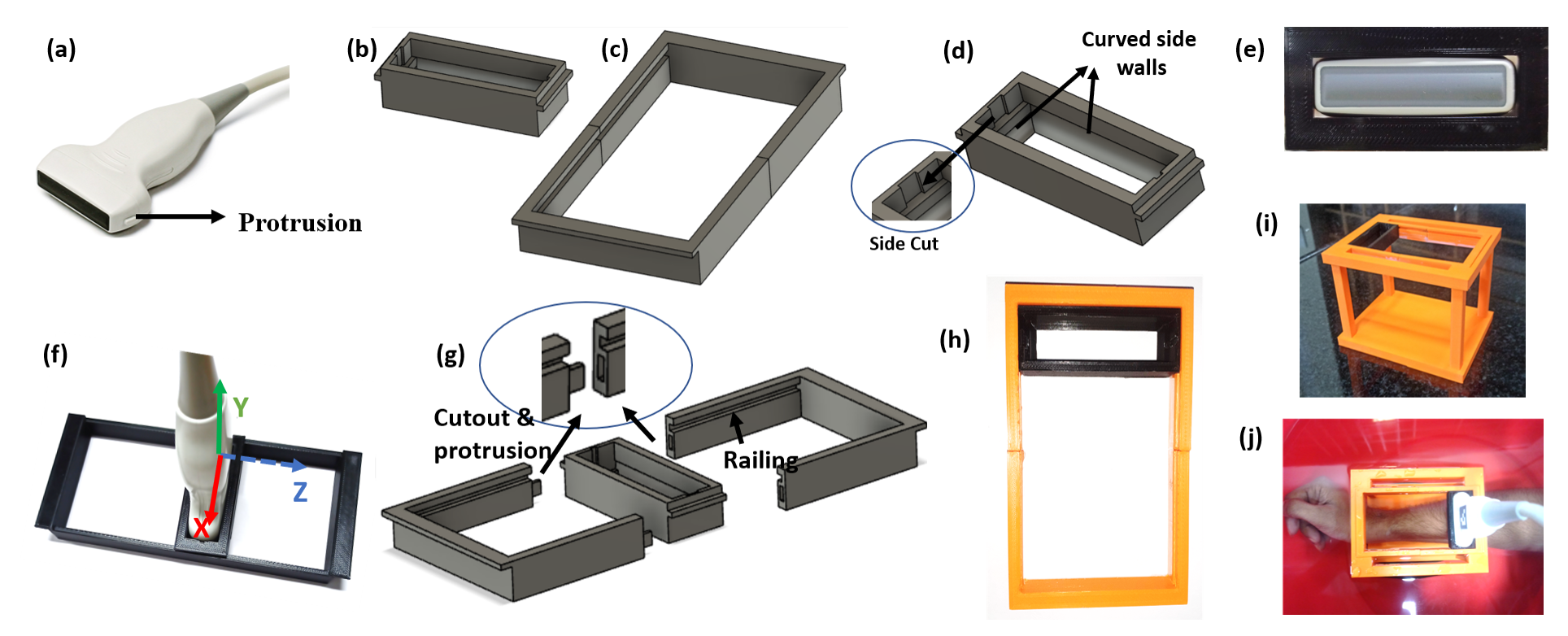}
    \caption[Mechanical track initial design]{(a) Verasonics L11-5v probe (b) probe mask (to be fitted to probe) (c) rectangular track (where the probe will be moved) (d) side cut and curved side walls of the mechanical track (e) \& (f) the probe on mechanical track (g) the novel mechanical track design (h) the 3D printed mechanical track (i) the stand for mounting the mechanical track (j) mechanical track and stand immersed in a water bath for data collection (in this case wrist scan)}
    \label{mechanical track full}
\end{figure*}

The mechanical track is designed with Fusion 360 to match the dimensions of the Verasonics L11-5v probe (Fig. \ref{mechanical track full} (a)) used for image acquisition in this work and can be designed for any probe in use. The mechanical track is designed as a modular block and consists of two components: the probe mask and the rectangular track (Fig. \ref{mechanical track full} (b) and (c)). The probe mask is a  structure that fits the probe precisely and has four curved side walls that match the probe geometry and improve vertical stability. Side cuts as shown in Fig. \ref{mechanical track full} (d), are incorporated into the mask to fit the probe's protrusions. The probe mask acts as a bounding box into which the probe is inserted (Fig. \ref{mechanical track full} (e)), effectively eliminating tilting issues along the x, y, and z-axes (as marked in Fig. \ref{mechanical track full} (f)). The rectangular track consists of a railing carved on its side walls on which the probe mask moves. It is made into different blocks with one end of the block having a cutout, while the other half has a protrusion such that nearby blocks can be joined together as shown in Fig. \ref{mechanical track full} (g) to extend the scanning length as desired. The mask is inserted between the two halves of the track, which is then joined to form a single block thereby creating a locking mechanism that prevents the vertical upward movement of the mask on the track. 

\begin{figure}[t]
    \centering
    \includegraphics[width=.3\textwidth]{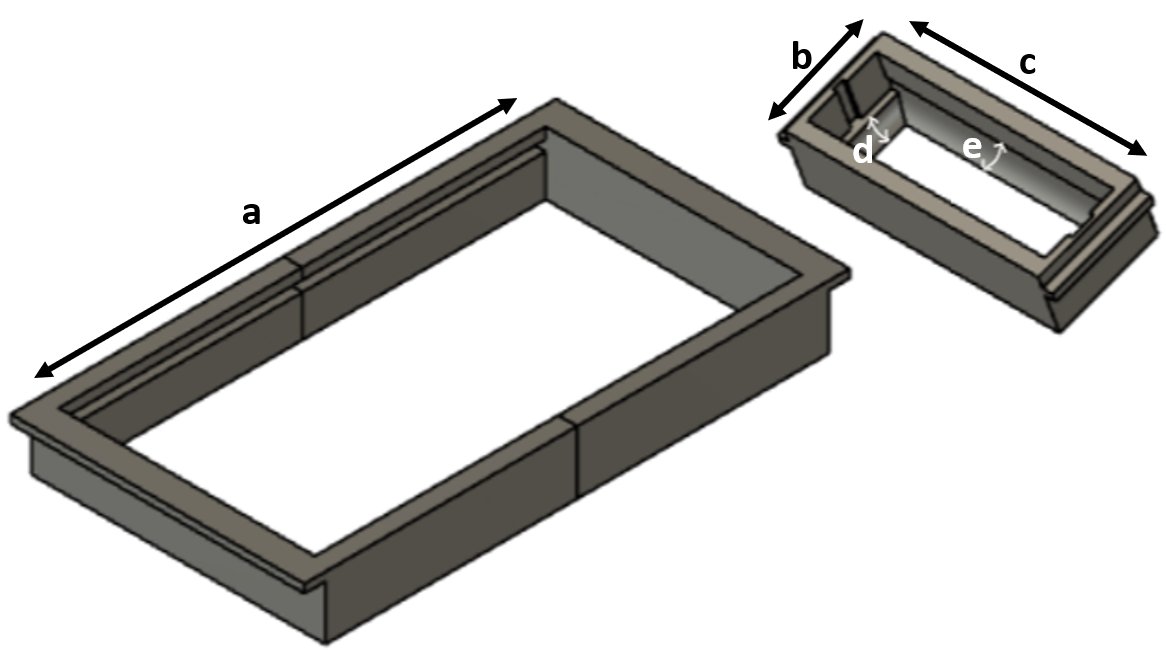}
    \caption{Mechanical track dimensions}
    \label{Mechanical Track dimensions}
\end{figure}

The mechanical track along with multi-angle plane wave-based high frame rate imaging \cite{high_framerate} ensures that the acquired frames are parallel and have a high correlation with the subsequent frames. This facilitates the use of simple linear interpolation between frames for generating the 3D volume when compared to other reconstruction techniques. The 3D printed mechanical track (Fig. \ref{mechanical track full} (h)) is mounted to a stand as shown in Fig. \ref{mechanical track full} (i) into which the scanning region is inserted and immersed in the water bath for image acquisition as shown in Fig. \ref{mechanical track full} (j). This setup effectively resolves the problems associated with tissue compression and patient discomfort typically experienced with conventional ultrasound systems, offering the advantages of a non-contact ultrasound system. Table \ref{tab:Rectangular Track Dimension} summarizes the dimensions of the mechanical track shown in Fig. \ref{Mechanical Track dimensions}.

\subsection{Probe Tracking using ArUco Marker}\label{sec: Probe tracking using ArUco marker}
\begin{figure}[t]
    \centering
    \includegraphics[width=.45\textwidth]{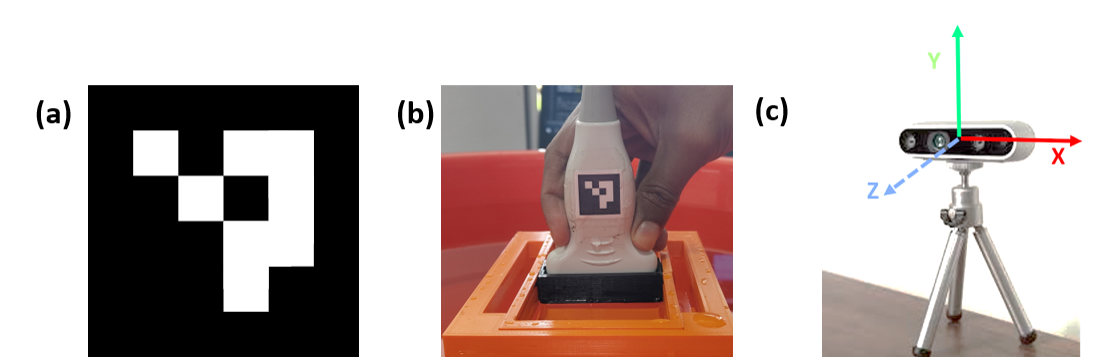}
    \caption{(a)ArUco Marker (b)ArUco Marker pasted on probe (c)Intel RealSense D435 camera} 
    \label{ArUco Mearker}
\end{figure}

An ArUco marker \cite{c31} which comprises a wide black boundary and a unique inner binary matrix as shown in Fig. \ref{ArUco Mearker} (a) is used for tracking the probe motion. The marker is pasted on the surface of the probe, as shown in Fig. \ref{ArUco Mearker} (b). The images of the probe with a marker is captured in real-time using  Intel RealSense depth camera (Fig. \ref{ArUco Mearker} (c)) and the position of the probe is estimated. 

\begin{table}[t]
    \centering
	\caption{Mechanical Track Dimensions}
	 \vspace*{5pt}
    \begin{tabular}{ccc}
    \hline
    \hline
    \textbf{Symbol}     &     \textbf{Dimension}    &     \textbf{Use} \\ 
    \textbf{in Fig. \ref{Mechanical Track dimensions}}     &     \textbf{(cm)}    &      \\ 
    
       \hline
    a - b  & 11.6 - 2.5 = 9.1 &  Length for linear motion of probe  \\ \hline
    c  & 5.3 & Width of the probe  \\ \hline
    d & Curve with radius of 7 &  For fitting the curved bottom part \\ \hline
    e & Curve with radius of 14 &  For fitting the curved bottom part\\ 
    \hline
    \hline
    \end{tabular}
    \label{tab:Rectangular Track Dimension}
\end{table}

The depth camera comprises high-definition RGB optical and infrared scanning depth sensors. The camera is initially calibrated using the MATLAB camera calibrator App as described in \cite{c32} to find the intrinsic matrix and the distortion parameters. 
The intrinsic matrix \textbf{(K)} and distortion coefficients \textbf{(d)} estimated for the Intel RealSense camera used in this work are as \eqref{Intrinsic Param of Realsense} and \eqref{Distortion Param of Realsense}.

\begin{equation}
\mathbf{K}=\left[\begin{array}{ccc}
422.451 & 0 & 427.466 \\
0 & 422.451 & 241.239 \\
0 & 0 & 1
\end{array}\right]
\label{Intrinsic Param of Realsense}
\end{equation}

\begin{equation}
\mathbf{d}=\left[\begin{array}{ccccc}
0.00690 & 0.8118 & 0.0 & 0.0 & -2.6234\\
\end{array}\right]
\label{Distortion Param of Realsense}
\end{equation}

The spatial location of the marker with respect to the camera coordinate frame is computed using a sequence of image processing steps applied to the captured RGB frame,  as proposed in \cite{c31}. Initially, edges within the captured image are detected, followed by the extraction of polygonal contours that could potentially be the markers. To validate a candidate as the required marker, its constituent bits are extracted and compared. For this, a perspective transformation is initially applied to attain the marker in its canonical form. Subsequently, Otsu's thresholding \cite{Otsu} is employed on the canonical image to separate the white and black bits. The image is divided into distinct cells based on the marker and the border sizes, and the counts of black and white pixels within each cell are estimated to determine whether a bit is white or black. The resulting bits are matched against the original pattern for verification. Upon identifying the marker, the spatial location of the ArUco marker's center in the $x$ and $y$ directions, relative to the camera's coordinate frame, is computed using the marker's corner pixels, the camera matrix, and distortion parameters. The $z$ location is then extracted from the depth direction captured by the Intel RealSense camera, corresponding to the marker's center in the frame.

\subsection{Data Acquisition}
 The ultrasound images are acquired with the mechanical track using the Verasonics Vantage 128 research ultrasound platform and the L11-5v linear array probe with 128 elements for multi-angle plane-wave transmission at a centre frequency of 7.6 MHz. The receive beamforming and image reconstruction are perfomed with Verasonics inbuilt delay and sum beamformer. The entire acquisition setup is shown in Fig. \ref{Verasonics} (a).

The real-time tracking data of the orientation of the probe is achieved by tracking the ArUco marker, and is paired with timestamps alongside the acquired ultrasound images. The Intel RealSense camera has a maximum frame rate of 60 fps, while the ultrasound images are acquired at a frame rate of 100 fps. The frames and orientation data are synchronized using timestamps.

\subsection{Estimation of Inter-Frame Spacing } \label{sec: Image spacing calculation }

\begin{figure*}[t]
    \centering
    \includegraphics[width=.95\textwidth]{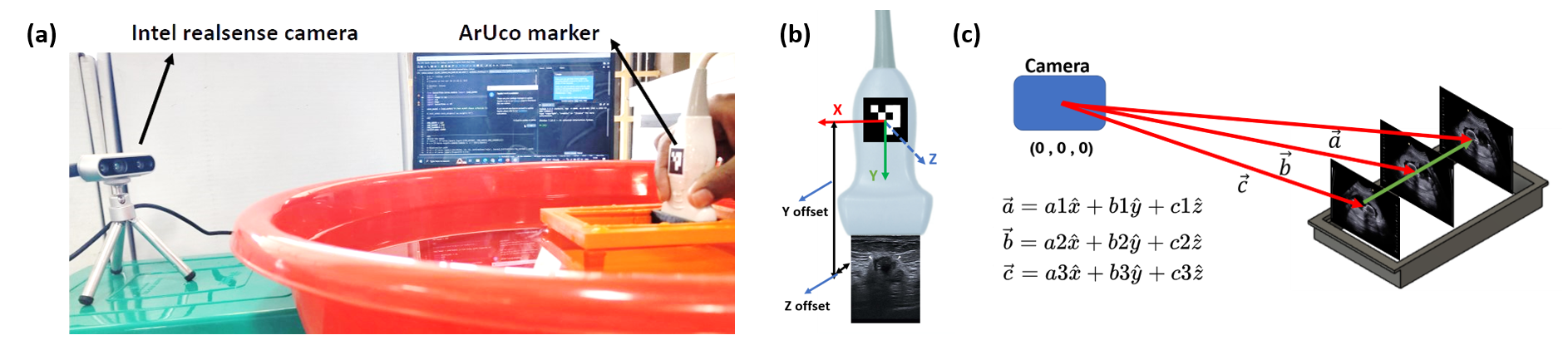}
    \caption{(a) Data acquisition setup consisting of the probe fitted on the mechanical track with scanning region immersed in water, the probe motion is captured by ArUco-based approach with the help of an Intel RealSense camera and high frame rate imaging employing the Verasonics Vantage 128 research ultrasound platform (b) ultrasound frame with respect to ArUco frame of reference (c) ultrasound images in the camera coordinate frame}
    \label{Verasonics}
\end{figure*}

The scanning region of interest remains at a fixed offset (in depth) from the probe, as shown in Fig. \ref{Verasonics} (b). Due to the placement of the ArUco marker, the acquired orientation data pertains to the probe which is mapped to the orientation of the ultrasound frames. As the 3D reconstruction relies on the relative motion between frames, the offsets can be disregarded which allows to obtain the orientation of the acquired ultrasound frames from the tracked position of the probe.

 \begin{figure}[t]
    \centering
    \includegraphics[width=.45\textwidth]{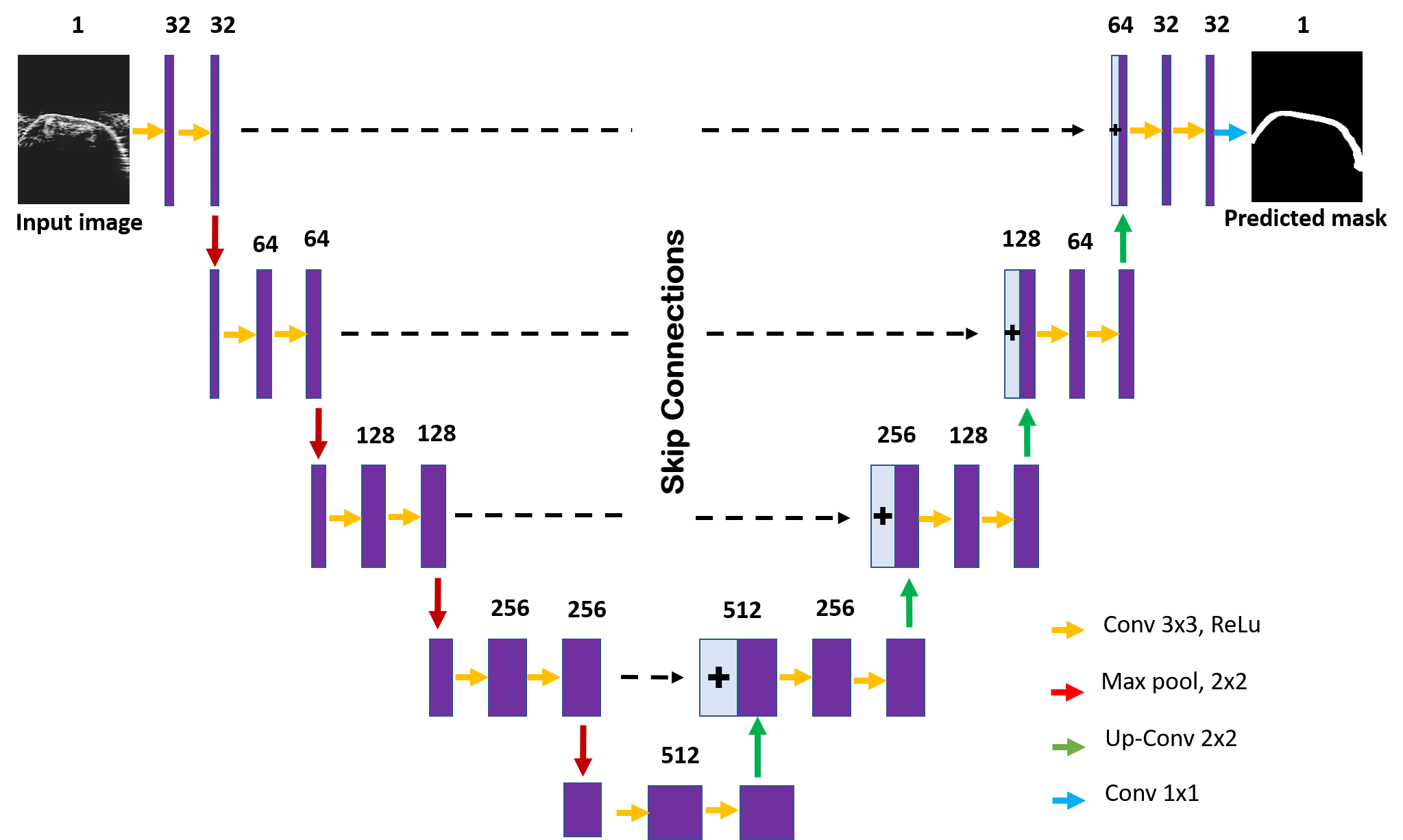}
    \caption{U-Net architecture used for image segmentation.}
    \label{Unet Architecture}
\end{figure}

\begin{figure*}[!t]
    \centering
    \includegraphics[width=1\textwidth]{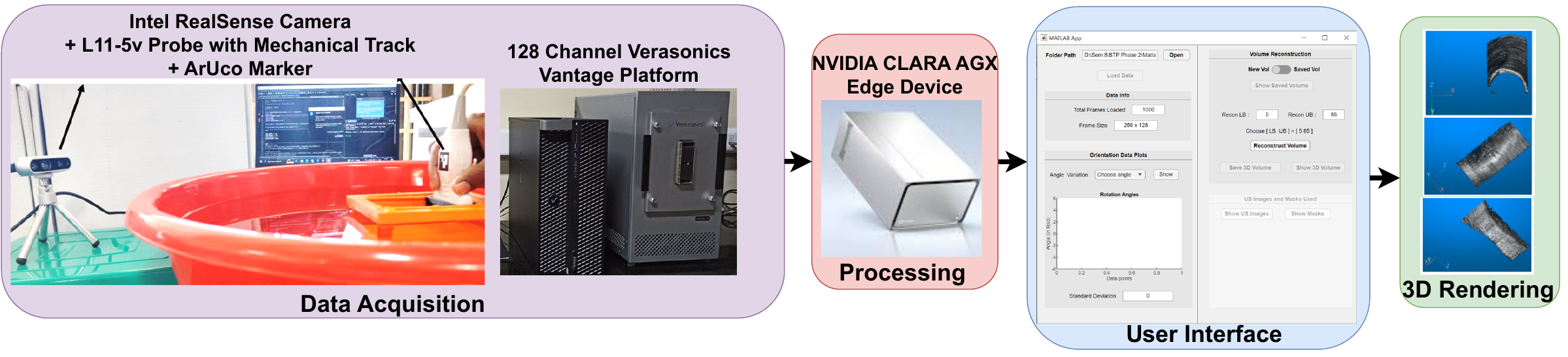}
    \caption{Setup for the proposed non contact 3D ultrasound pipeline. The data acquisition is done using the Verasonics 128-channel research platform employing the L11-5v probe fitted onto the mechanical track. A combination of Intel RealSense camera and ArUco marker is employed to track the probe motion. The acquired images were processed to construct the volume using Nvidia Clara AGX with almost $6$ times speedup. The volume reconstructed was visualised using the Volume Viewer tool in Matlab}
    \label{hardware implementation}
\end{figure*}

\AddToShipoutPictureBG*{%
  \AtPageUpperLeft{%
    \setlength\unitlength{1in}%
    \hspace*{\dimexpr0.5\paperwidth\relax}
    \makebox(0,-0.75)[c]{\textcolor{red}{\large This is an originally submitted version and has not been reviewed by independent peers.}}%
}}

\AddToShipoutPictureBG*{%
  \AtPageLowerLeft{%
    \setlength\unitlength{1in}%
    \hspace*{\dimexpr0.5\paperwidth\relax}
    \makebox(0,0.75)[c]{\textcolor{red}{\textit{This work is licensed under a \href{https://creativecommons.org/licenses/by-nc-nd/4.0/}{Creative Commons Attribution-NonCommercial-NoDerivatives (CC-BY-NC-ND) 4.0 License.}}}}%
}}

\begin{figure*}[t]
    \centering
    \includegraphics[width=1\textwidth]{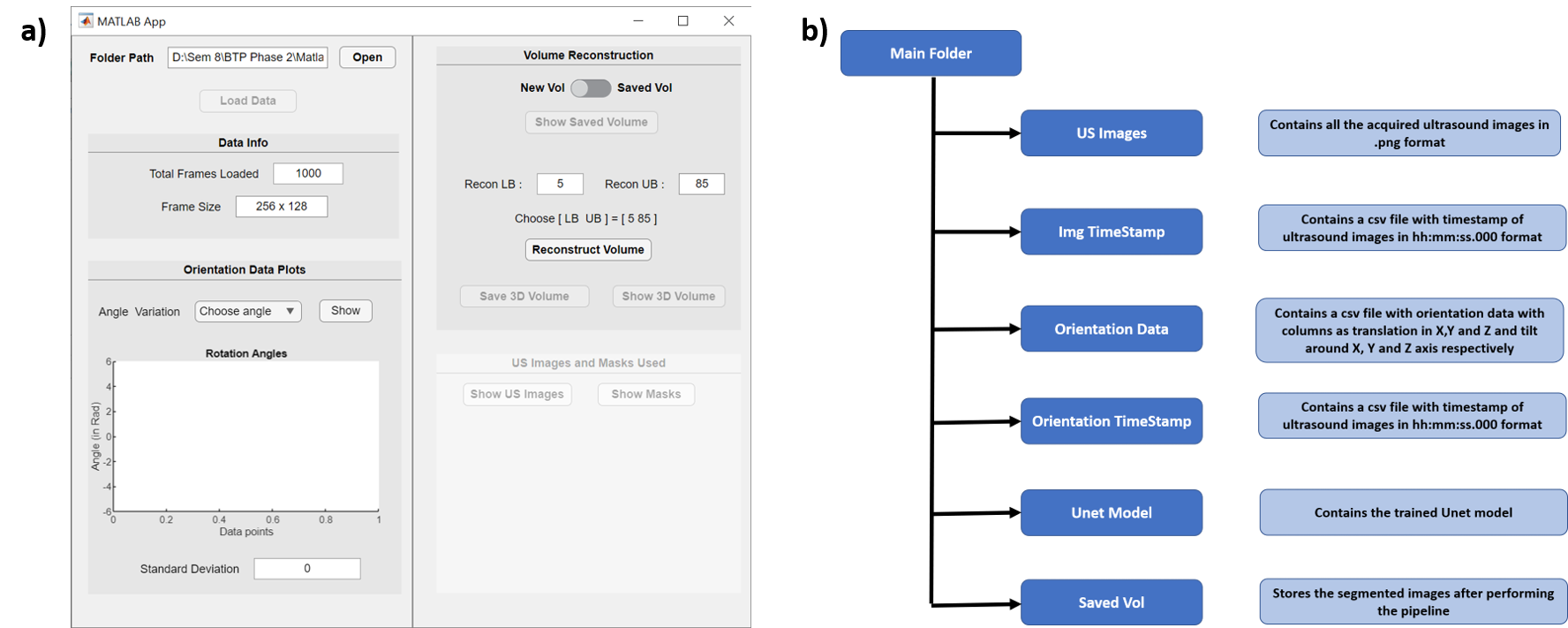}
    \caption{ a) User Interface of the standalone MATLAB application employed for 3D reconstruction b) The folder hierarchy followed in the proposed stand-alone MATLAB application} 
    \label{UI}
\end{figure*}



\begin{figure*}[t]
    \centering
    \includegraphics[width=.8\textwidth]{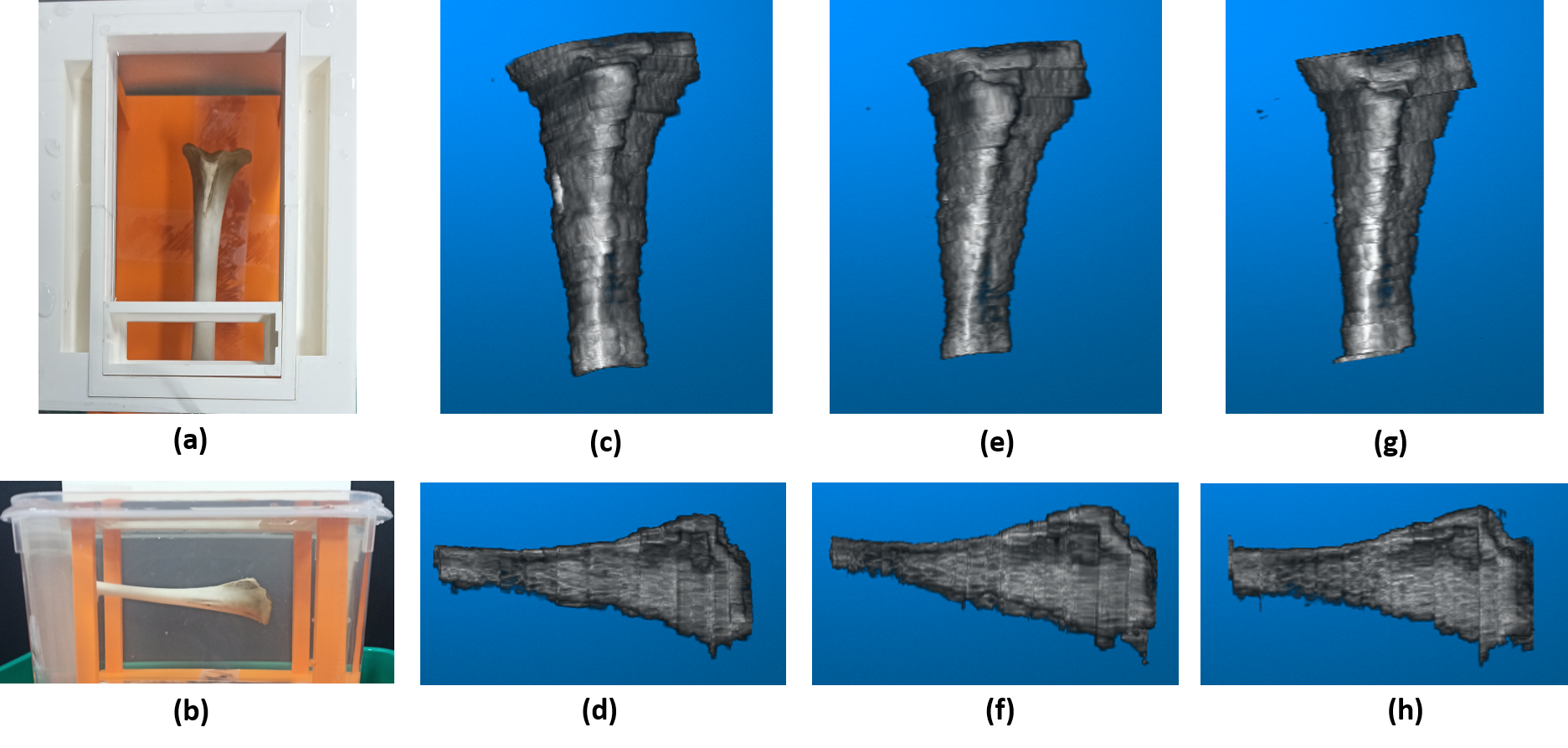}
    \caption{Ex-vivo experiment results (a) \& (b) Tibia of goat placed in the water bath for the experiment, (c) \& (d) reconstructed volume in experiment 1, (e) \& (f) reconstructed volume in experiment 2, (g) \& (h) reconstructed volume in experiment 3}
    \label{Exvivo}
\end{figure*}

The distance of the centre of the ultrasound frame from the camera is as shown in Fig. \ref{Verasonics} (c), where $\vec{\mathbf{a}}$, $\vec{\mathbf{b}}$, $\vec{\mathbf{c}}$ represents the distance vectors of the frames from the camera. 
\begin{equation}
\begin{aligned}
 &\vec{a}=a 1 \hat{x}+b 1 \hat{y}+c 1 \hat{z} \\
 &\vec{b}=a 2 \hat{x}+b 2 \hat{y}+c 2 \hat{z} \\
 &\vec{c}=a 3 \hat{x}+b 3 \hat{y}+c 3 \hat{z}  
\end{aligned}
\end{equation}

The displacements in each axis ($a1$,$b1$,$c1$) is obtained by tracking the ArUco marker. Due to the motion of the probe on the mechanical track, the acquired ultrasound frames are parallel to each other and therefore the inter-frame spacing can be estimated as the norm of the difference of their position vectors from the camera. For example, the spacing between the first and second frames and the first and third frames shown in Fig. \ref{Verasonics} (c) can be found as \eqref{eqn_interframe_spacing_1} and \eqref{eqn_interframe_spacing_2},
\begin{eqnarray}
f_{ab} = \lVert \vec{b} -  \vec{a} \rVert  \label{eqn_interframe_spacing_1}\\
    f_{ac} = \lVert \vec{c} -  \vec{a} \rVert \label{eqn_interframe_spacing_2}
\end{eqnarray}

\AddToShipoutPictureBG*{%
  \AtPageUpperLeft{%
    \setlength\unitlength{1in}%
    \hspace*{\dimexpr0.5\paperwidth\relax}
    \makebox(0,-0.75)[c]{\textcolor{red}{\large This is an originally submitted version and has not been reviewed by independent peers.}}%
}}

\AddToShipoutPictureBG*{%
  \AtPageLowerLeft{%
    \setlength\unitlength{1in}%
    \hspace*{\dimexpr0.5\paperwidth\relax}
    \makebox(0,0.75)[c]{\textcolor{red}{\textit{This work is licensed under a \href{https://creativecommons.org/licenses/by-nc-nd/4.0/}{Creative Commons Attribution-NonCommercial-NoDerivatives (CC-BY-NC-ND) 4.0 License.}}}}%
}}

\subsection{Pre-processing of Image Frames}
\label{sec:Data Pre-processing}
Ultrasound images are often characterized by speckles, which complicate the segmentation of the images and visualization of the reconstructed 3D volume. To enhance the accuracy of image segmentation, the acquired images are subjected to a series of image processing operations outlined below.

\subsubsection{Log Compression}
Log compression enhances the low-intensity values in images with strong reflections. 

\subsubsection{Median filtering}
The salt and pepper noise in the log compressed images are suppressed with a non-linear median filter with a $3\times3$ overlapping window.

\subsubsection{Thresholding}
Pixel values are confined within a predefined range determined by empirical maximum and minimum values. Any pixel values that fall outside this range are adjusted to align with these boundary values. 

\subsubsection{Contrast Limited Adaptive Histogram Equalization}
The image quality is further enhanced by contrast limited adaptive histogram equalisation (CLAHE) \cite{c10}. It is a variant of adaptive histogram equalisation that distributes the part of the histogram that exceeds the specified limit equally across all histograms to constrain the contrast enhancement to reduce the amplified noise.

\subsection{Region Segmentation}
\label{sec: Segmentation}

The segmentation of the wrist bone is achieved by adopting the U-Net based image segmentation \cite{c33} shown in Fig. \ref{Unet Architecture}. The input images are reshaped into dimensions of  $256\times128$ before feeding to the network. The architecture consists of a contracting path on the left and an expansive path on the right. The contracting path applies a sequence of two $3\times3$ convolutions (with the same padding), followed by a rectified linear unit (ReLU) and a 2x2 max pooling operation with stride 2 for downsampling at each stage. At each downsampling step, the number of feature channels is doubled. The expansive path involves an upsampling of the feature map, followed by a 2x2 up-convolution that halves the number of feature channels. The resulting feature map is concatenated with the corresponding feature map from the contracting path using skip connections, which help the network retain high-resolution information from the input image and improve the segmentation accuracy. The segmentation mask, which represents pixel-wise classification, is obtained by passing the output of the last expansion layer through a 1x1 convolution with a sigmoid activation function. The segmentation mask obtained from the U-Net model is multiplied with the corresponding images to get the segmented images for 3D reconstruction.

\subsection{Volume Reconstruction and Visualization}
The segmented ultrasound images are employed for the volume reconstruction. The frames are first rearranged in a monotonically increasing fashion based on the estimated image spacing (as explained in the previous subsection) and the reconstruction is performed using the conventional linear interpolation.The number of intermediate frames between each acquired frame is determined based on inter-frame spacing such that the spacing between all the frames is 0.1 mm  after interpolation. Since the acquired frames are parallel to each other due to the use of mechanical track and a high correlation exists between the frames due to high frame rate, linear interpolation  simplifies the volume reconstruction and reduce the computational complexity. The reconstructed volume is visualized with volume rendering available with the Volume Viewer in MATLAB\textsuperscript{\textregistered}.

\AddToShipoutPictureBG*{%
  \AtPageUpperLeft{%
    \setlength\unitlength{1in}%
    \hspace*{\dimexpr0.5\paperwidth\relax}
    \makebox(0,-0.75)[c]{\textcolor{red}{\large This is an originally submitted version and has not been reviewed by independent peers.}}%
}}

\AddToShipoutPictureBG*{%
  \AtPageLowerLeft{%
    \setlength\unitlength{1in}%
    \hspace*{\dimexpr0.5\paperwidth\relax}
    \makebox(0,0.75)[c]{\textcolor{red}{\textit{This work is licensed under a \href{https://creativecommons.org/licenses/by-nc-nd/4.0/}{Creative Commons Attribution-NonCommercial-NoDerivatives (CC-BY-NC-ND) 4.0 License.}}}}%
}}

\section{Experimental Setup}
\label{EXPERIMENTS AND RESULTS}

\begin{figure*}[t]
    \centering
    \includegraphics[width=1\textwidth]{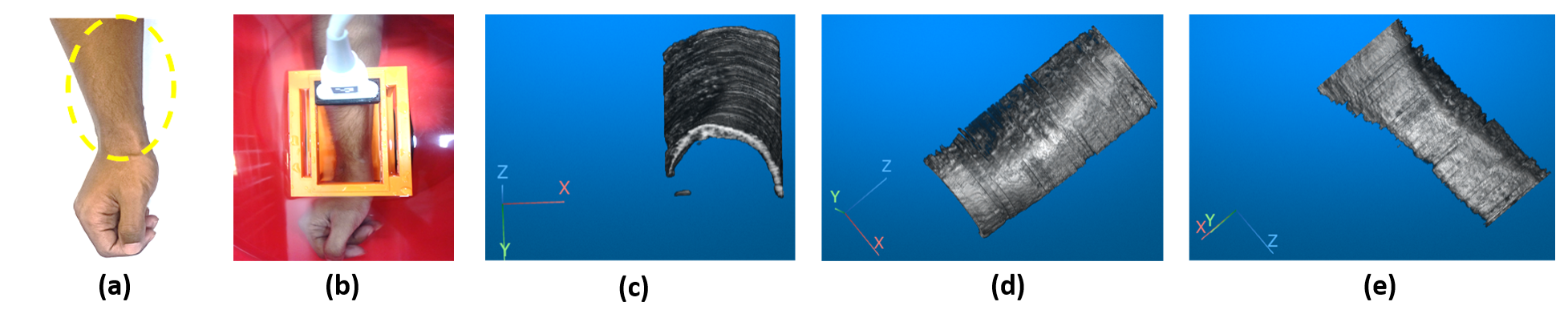}
    \caption[Radial bone scanning position] {(a) Radial bone in human forearm (b) scanning position of the forearm. (c), (d) \& (e) different views of the reconstructed volume of radius bone}
    \label{Radius bone}
\end{figure*}
\begin{figure*}[t]
    \centering
    \includegraphics[width=1\textwidth]{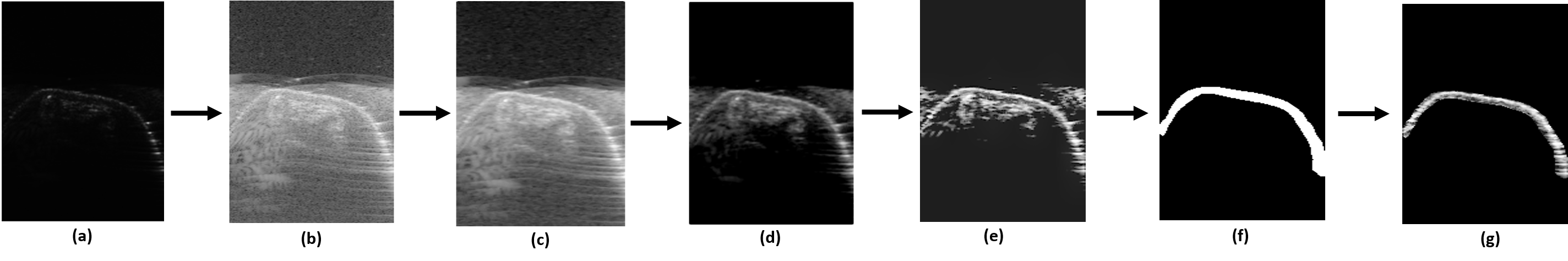}
    \caption[Image pre-processing steps for \textit{in-vivo} experiment]{Image pre-processing steps for \textit{in-vivo} radial bone experiment (a) acquired ultrasound image (b) log compressed image (c) median filtered image (d) thresholded image (e) contrast limited adaptive histogram equalized (CLAHE) image (f) segmentation mask (g) segmented image}
    \label{Image Enhancing Steps}
\end{figure*}

\begin{figure}[t]
    \centering
    \includegraphics[width=.47\textwidth]{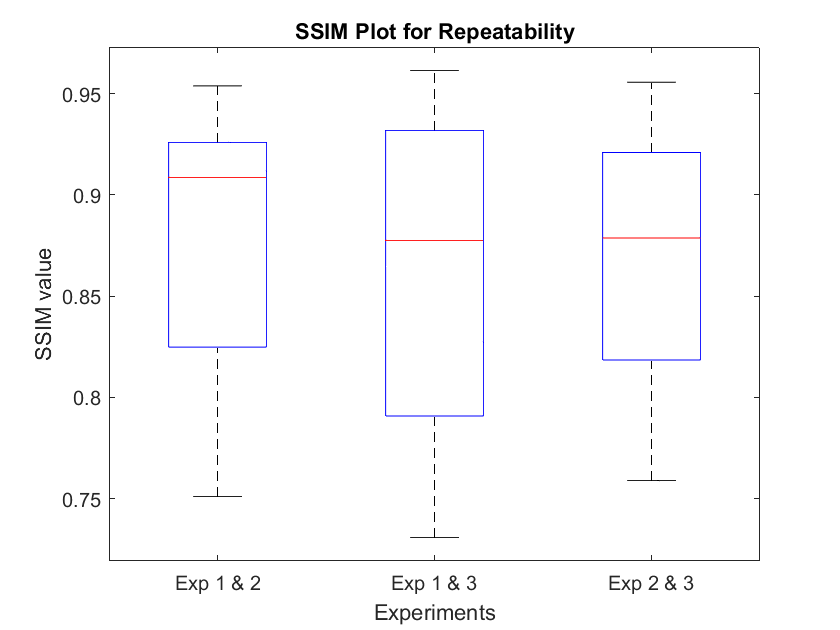}
    \caption{Box plot comparing the SSIM values among the corresponding frames in each of the three experiments conducted for 3D reconstruction of bone}
    \label{SSIM}
\end{figure}

\subsection{Implementation Details}
The proposed framework as shown in Fig. \ref{hardware implementation} consists of a) the image acquisition done using the 128 channel Verasonics Vantage platform equipped with L11-5v linear array transducer probe fitted on the proposed mechanical track, b) the position data was acquired using the ArUco marker-based tracking system using the Intel RealSense camera and c) the acquired data were processed to generate the volumetric data using the NVIDIA Clara AGX. The execution time for data processing got reduced to 18 secs on NVIDIA Clara System compared to 100 secs on a CPU with  Intel(R) Core(TM) i5-1035G1 @ 1.00GHz, 1190 Mhz and 4 Core(s) resulting in almost $6$ times reduction in the computational time. 
\subsection{Standalone Matlab Application}
A stand-alone MATLAB application has been developed to improve the usability and accessibility of the developed 3D reconstruction pipeline. This application provides a user-friendly interface for the users to input the ultrasound images and orientation data collected and reconstruct the 3D volume from the input data. \\
The user interface of the standalone application as is shown in Fig. \ref{UI} (a). For the reconstruction of the 3D volume, the MATLAB application must be loaded with a primary directory comprising six subfolders, each containing the required data. The hierarchy and the contents of the loaded directory are shown in Fig. \ref{UI} (b). The details and the demo of the MATLAB application can be seen at the \href{https://youtu.be/_Btqi7lQSgw}{link}.

\subsection{\textit{Ex-vivo} Bone Phantom Study}
The proposed approach is validated with an \textit{ex-vivo} experiment using a bone (goat tibia) placed in a water bath as shown in Fig. \ref{Exvivo} (a) and (b). The beamformed ultrasound images with dimensions of $256\times190$ pixels and their corresponding orientations are acquired as proposed. The images are pre-processed, bone segmented and arranged based on image spacing to generate the 3D volume. The U-Net model for segmentation is trained for 100 images of the bone and achieved an F1 score of 0.85. The experiment is repeated thrice to ensure repeatability. 

\AddToShipoutPictureBG*{%
  \AtPageUpperLeft{%
    \setlength\unitlength{1in}%
    \hspace*{\dimexpr0.5\paperwidth\relax}
    \makebox(0,-0.75)[c]{\textcolor{red}{\large This is an originally submitted version and has not been reviewed by independent peers.}}%
}}

\AddToShipoutPictureBG*{%
  \AtPageLowerLeft{%
    \setlength\unitlength{1in}%
    \hspace*{\dimexpr0.5\paperwidth\relax}
    \makebox(0,0.75)[c]{\textcolor{red}{\textit{This work is licensed under a \href{https://creativecommons.org/licenses/by-nc-nd/4.0/}{Creative Commons Attribution-NonCommercial-NoDerivatives (CC-BY-NC-ND) 4.0 License.}}}}%
}}

\begin{table}[t]
    \centering
	\caption{Standard deviation of tilt around X, Y, and Z axis in radians}
	 \vspace*{5pt}
    \begin{tabular}{cccc}
    \hline
    \hline
    \textbf{Experiment No}      &       \textbf{X axis}     &       \textbf{Y axis}     &       \textbf{Z axis}\\ 
       \hline
    1  & 0.082 & 0.072 & 0.213  \\ 
    2  & 0.099 & 0.043 & 0.220 \\ 
    3  & 0.098  & 0.034 & 0.171 \\ 
    \hline
    \hline
    \end{tabular}
    \label{tab:Standard deviation of tilt}
\end{table}

\AddToShipoutPictureBG*{%
  \AtPageUpperLeft{%
    \setlength\unitlength{1in}%
    \hspace*{\dimexpr0.5\paperwidth\relax}
    \makebox(0,-0.75)[c]{\textcolor{red}{\large This is an originally submitted version and has not been reviewed by independent peers.}}%
}}

\AddToShipoutPictureBG*{%
  \AtPageLowerLeft{%
    \setlength\unitlength{1in}%
    \hspace*{\dimexpr0.5\paperwidth\relax}
    \makebox(0,0.75)[c]{\textcolor{red}{\textit{This work is licensed under a \href{https://creativecommons.org/licenses/by-nc-nd/4.0/}{Creative Commons Attribution-NonCommercial-NoDerivatives (CC-BY-NC-ND) 4.0 License.}}}}%
}}

\subsection{\textit{In-vivo} Radial bone Study}
The proposed approach is validated with an \textit{in-vivo} experiment with the radial bone of the human forearm (Fig. \ref{Radius bone} (a)). Fig. \ref{Radius bone} (a) and (b) illustrate the scanning position of the forearm in the \textit{in-vivo} experiment. The quality of the acquired ultrasound images is improved using the data pre-processing. A U-Net model trained with 100 ultrasound images of the cross-sectional views of the radius bone of multiple healthy humans was used to predict the segmentation masks. The U-Net model achieved an F1-score of 0.8241.

\section{Results}
Fig. \ref{Exvivo} (c) and (d), (e) and (f), and (g) and (h) show the reconstructed bone images from experiments 1, 2, and 3 respectively. The structural similarity index (SSIM) value for the corresponding frames in each of the three experiments is computed and shown in the box plot in Fig. \ref{SSIM}. A high median value is observed in the plot which confirms that the proposed framework ensures repeatability in acquisition. Table \ref{tab:Standard deviation of tilt} summarises standard deviations in the angles probe made with the x, y, and z axis when moved on the mechanical track with respect to the camera coordinate frame of reference (Fig. \ref{ArUco Mearker} (c)). From the table, we can infer that the use of a mechanical track for data acquisition helps to reduce the tilting of the probe and acquire parallel frames, resulting in low complex reconstruction approaches.

Fig. \ref{Radius bone} (c), (d), and (e) shows the multiple angle views of the reconstructed radial bone from the \textit{in-vivo} study. Fig. \ref{Image Enhancing Steps} shows the image pre-processing steps for the experiment. The acquired frames (as shown in Fig. \ref{Image Enhancing Steps} (a)) are log compressed (as shown in Fig. \ref{Image Enhancing Steps} (b)) and median filtered (as shown in Fig. \ref{Image Enhancing Steps} (c)) to improve on the visual quality. The enhanced image is thresholded empirically (as shown in Fig. \ref{Image Enhancing Steps} (d)) and the CLAHE algorithm is applied as shown in Fig. \ref{Image Enhancing Steps} (e). The segmentation mask (as shown in Fig. \ref{Image Enhancing Steps} (f)) is applied to extract the bone region of interest as shown in Fig. \ref{Image Enhancing Steps} (g). Once the bone region is extracted a 3D reconstruction as in Fig. \ref{Radius bone} (c)-(e) can be obtained.

\section{Discussion and Conclusion}

In this work, a novel pipeline for non-contact freehand scanning-based 3D ultrasound imaging is proposed. The proposed approach comprises of 
\begin{itemize}
    \item a 3D-printed tilt-resistant linear mechanical track which ensures minimal tilt and linear scans.
    \item a high frame rate ultrasound image acquisition based on multi-angle plane wave transmit and compounding at receive, which leverages the redundancy across frames for enhanced image quality and low complex reconstruction.
    \item an ArUco tracker for probe motion estimation which ensures the reconstruction considers the speed of the scan making it operator-independent.
    \item an accelerated pre-processing and 3D reconstruction using the edge compute device (NVIDIA CLARA AGX).
    \item a MATLAB-based user interface which provides an interactive environment to the user.    
\end{itemize}
The experiments demonstrated that the use of the mechanical track reduced errors due to probe slippage or tilt during scanning. Table \ref{tab:Standard deviation of tilt} validates our assumption of planar motion (constant rotation angle/ tilt around all axes) on the mechanical track. Moreover, the high frame rate acquisition resulted in a high correlation among subsequent frames, which in combination, resulted in a cost-effective and simple 3D reconstruction pipeline (e.g. linear interpolation will suffice). It has to be noted that the cost of electromagnetic and optical tracker-based sensors for position tracking was above \$1000 while the proposed system costs around \$300 only and also enables non-contact scanning.  
The proposed pipeline finds potential applications in pediatric ultrasound imaging where X-ray imaging can be potentially dangerous due to ionising radiation. In such cases, the proposed approach can be used to identify fractures and their location and can be employed for periodic scanning during the recovery phase. Also, since the proposed approach is non-contact, direct pressure on the fracture/wounded regions is avoided reducing the discomfort caused to the subjects. 

However, there are some limitations to the proposed approach which will be addressed in future iterations. Although the reconstructed structure appeared almost identical to the actual bone, the accuracy of the reconstructed volume when compared to the existing techniques needs to be evaluated. Further, the design of the mechanical track shall be extended to incorporate the scanning of tissue structures of different shapes in future work. In this work, water is employed as the couplant, but this may not be feasible in all scenarios and hence future work involves the 3D design of the scanning system similar to a computed tomography (CT) scan.

\AddToShipoutPictureBG*{%
  \AtPageUpperLeft{%
    \setlength\unitlength{1in}%
    \hspace*{\dimexpr0.5\paperwidth\relax}
    \makebox(0,-0.75)[c]{\textcolor{red}{\large This is an originally submitted version and has not been reviewed by independent peers.}}%
}}

\AddToShipoutPictureBG*{%
  \AtPageLowerLeft{%
    \setlength\unitlength{1in}%
    \hspace*{\dimexpr0.5\paperwidth\relax}
    \makebox(0,0.75)[c]{\textcolor{red}{\textit{This work is licensed under a \href{https://creativecommons.org/licenses/by-nc-nd/4.0/}{Creative Commons Attribution-NonCommercial-NoDerivatives (CC-BY-NC-ND) 4.0 License.}}}}%
}}

\AddToShipoutPictureBG*{%
  \AtPageUpperLeft{%
    \setlength\unitlength{1in}%
    \hspace*{\dimexpr0.5\paperwidth\relax}
    \makebox(0,-0.75)[c]{\textcolor{red}{\large This is an originally submitted version and has not been reviewed by independent peers.}}%
}}

\AddToShipoutPictureBG*{%
  \AtPageLowerLeft{%
    \setlength\unitlength{1in}%
    \hspace*{\dimexpr0.5\paperwidth\relax}
    \makebox(0,0.75)[c]{\textcolor{red}{\textit{This work is licensed under a \href{https://creativecommons.org/licenses/by-nc-nd/4.0/}{Creative Commons Attribution-NonCommercial-NoDerivatives (CC-BY-NC-ND) 4.0 License.}}}}%
}}


\begin{thebibliography}{99}
\bibliographystyle{ieeetr}
\bibitem{c1} Mohamed, Farhan, and C. Vei Siang. "A survey on 3D ultrasound reconstruction techniques." Artificial Intelligence—Applications in Medicine and Biology (2019): 73-92.

\bibitem{c30} Zhang, Xiang, et al. "Full noncontact laser ultrasound: first human data." Light: Science \& Applications 8.1 (2019): 119.

\bibitem{c14} Wen, Tiexiang, et al. "GPU-based volume reconstruction for freehand 3D ultrasound imaging." 39th Annual International Conference of the IEEE Engineering in Medicine and Biology Society. IEEE, 2017

\bibitem{c16} Woo, Jeongdong, and Yongrae Roh. "Ultrasonic 2D matrix array transducer for volumetric imaging in real-time." 2012 ieee international ultrasonics symposium. IEEE, 2012.


\bibitem{c50}Smith, Stephen W., et al. "Two dimensional arrays for 3-D ultrasound imaging." 2002 IEEE Ultrasonics Symposium, 2002. Proceedings.. Vol. 2. IEEE, 2002.

\bibitem{c2} Huang, Qinghua, and Zhaozheng Zeng. "A review on real-time 3D ultrasound imaging technology." BioMed research international 2017.

\bibitem{c52}Huang, Qinghua, Jiulong Lan, and Xuelong Li. "Robotic arm based automatic ultrasound scanning for three-dimensional imaging." IEEE Transactions on Industrial Informatics 15.2 (2018): 1173-1182.

\bibitem{c18} Kaminski, Jakub T., Khashayar Rafatzand, and Haichong K. Zhang. "Feasibility of robot-assisted ultrasound imaging with force feedback for assessment of thyroid diseases." Medical Imaging 2020: Image-Guided Procedures, Robotic Interventions, and Modeling. Vol. 11315. SPIE, 2020.

\bibitem{c3} Daoud, Mohammad I., et al. "Freehand 3D ultrasound imaging system using electromagnetic tracking." 2015 International Conference on Open Source Software Computing (OSSCOM). IEEE, 2015.

\bibitem{c45}MacGillivray, Thomas J., et al. "3D freehand ultrasound for in vivo determination of human skeletal muscle volume." Ultrasound in medicine \& biology 35.6 (2009): 928-935.

\bibitem{c44}Chung, Shao-Wen, Cho-Chiang Shih, and Chih-Chung Huang. "Freehand three-dimensional ultrasound imaging of carotid artery using motion tracking technology." Ultrasonics 74 (2017): 11-20

\bibitem{c40} Prevost, Raphael, et al. "3D freehand ultrasound without external tracking using deep learning." Medical image analysis 48 (2018): 187-202

 \bibitem{c41}Guo, Hengtao, et al. "Sensorless freehand 3D ultrasound reconstruction via deep contextual learning." Medical Image Computing and Computer Assisted Intervention–MICCAI 2020: 23rd International Conference, Lima, Peru, October 4–8, 2020, Proceedings, Part III 23. Springer International Publishing, 2020.

  \bibitem{c48} Guo, Hengtao, et al. "Ultrasound Volume Reconstruction From Freehand Scans Without Tracking." IEEE Transactions on Biomedical Engineering 70.3 (2022): 970-979.


\bibitem{c47} Chang, Ruey-Feng, et al. "3-D US frame positioning using speckle decorrelation and image registration." Ultrasound in medicine \& biology 29.6 (2003): 801-812.

\bibitem{c49}Laporte, Catherine, and Tal Arbel. "Probabilistic speckle decorrelation for 3d ultrasound." Medical Image Computing and Computer-Assisted Intervention–MICCAI 2007: 10th International Conference, Brisbane, Australia, October 29-November 2, 2007, Proceedings, Part I 10. Springer Berlin Heidelberg, 2007.

\bibitem{c42} Gao, Haitao, et al. "Wireless and sensorless 3D ultrasound imaging." Neurocomputing 195 (2016): 159-171.

\bibitem{c43} Hafizah, Mahani, Tan Kok, and Eko Supriyanto. "Development of 3D image reconstruction based on untracked 2D fetal phantom ultrasound images using VTK." WSEAS transactions on signal processing 6.4 (2010): 145-154.


\bibitem{c35} Solberg, Ole Vegard, et al. "Freehand 3D ultrasound reconstruction algorithms—a review." Ultrasound in medicine \& biology 33.7 (2007): 991-1009.


\bibitem{c36} Chen, Xiankang, et al. "Reconstruction of freehand 3D ultrasound based on kernel regression." Biomedical engineering online 13.1 (2014): 1-15.


\bibitem{c37} Gobbi, David G., and Terry M. Peters. "Interactive intra-operative 3D ultrasound reconstruction and visualization." Medical Image Computing and Computer-Assisted Intervention—MICCAI 2002: 5th International Conference Tokyo, Japan, September 25–28, 2002 Proceedings, Part II 5. Springer Berlin Heidelberg, 2002.

\bibitem{c60} Sanches, Joao M., and Jorge S. Marques. "A multiscale algorithm for three-dimensional free-hand ultrasound." Ultrasound in medicine \& biology 28.8 (2002): 1029-1040.

\bibitem{c61} Huang, Qinghua, et al. "Bezier interpolation for 3-D freehand ultrasound." IEEE Transactions on Human-Machine Systems 45.3 (2014): 385-392.

\bibitem{c38}Lindseth, Frank, et al. "Ultrasound-based guidance and therapy." Advancements and breakthroughs in ultrasound imaging. IntechOpen, 2013.

\bibitem{c62} Gobbi, David G., and Terry M. Peters. "Interactive intra-operative 3D ultrasound reconstruction and visualization." International conference on medical image computing and computer-assisted intervention. Berlin, Heidelberg: Springer Berlin Heidelberg, 2002.

\bibitem{c64}Zhang, Qi, Roy Eagleson, and Terry M. Peters. "Volume visualization: a technical overview with a focus on medical applications." Journal of digital imaging 24 (2011): 640-664.

\bibitem{c63} Tan, Jianhao, et al. "Design of 3D visualization system based on VTK utilizing marching cubes and ray casting algorithm." 2016 8th International Conference on Intelligent Human-Machine Systems and Cybernetics (IHMSC). Vol. 2. IEEE, 2016.
  
\bibitem{confpaper} Antony Jerald, Madhavanunni A. N., Gayathri Malamal, Pisharody Harikrishnan Gopalakrishnan, and Mahesh Raveendranatha Panicker, "A Simplified 3D Ultrasound Freehand Imaging Framework Using 1D Linear Probe and Low-Cost Mechanical Track" accepted in 8th International Conference on Computer Vision \& Image Processing (CVIP-2023), Nov 2023.

\bibitem{high_framerate}Bercoff, Jeremy. "Ultrafast ultrasound imaging." Ultrasound imaging-Medical applications (2011): 3-24.

\bibitem{c31}Garrido-Jurado, Sergio, et al. "Automatic generation and detection of highly reliable fiducial markers under occlusion." Pattern Recognition 47.6 (2014): 2280-2292.

\bibitem{c32} Fetić, Azra, Davor Jurić, and Dinko Osmanković. "The procedure of a camera calibration using Camera Calibration Toolbox for MATLAB." 2012 Proceedings of the 35th International Convention MIPRO. IEEE, 2012.

\bibitem{Otsu}N. Otsu, "A Threshold Selection Method from Gray-Level Histograms," in IEEE Transactions on Systems, Man, and Cybernetics, vol. 9, no. 1, pp. 62-66, Jan. 1979, doi: 10.1109/TSMC.1979.4310076.

\bibitem{c10} Zuiderveld, Karel J.. “Contrast Limited Adaptive Histogram Equalization.” Graphics gems (1994).

\bibitem{c33} Ronneberger, Olaf, Philipp Fischer, and Thomas Brox. "U-net: Convolutional networks for biomedical image segmentation." Medical Image Computing and Computer-Assisted Intervention–MICCAI 2015: 18th International Conference, Munich, Germany, October 5-9, 2015, Proceedings, Part III 18. Springer International Publishing, 2015.

\end{thebibliography}
\end{document}